\documentstyle[prl,aps]{revtex}  
\begin{document}
\draft

\title{ NON-EXTENSIVE STATISTICS AND SOLAR NEUTRINOS }
\author{ G. Kaniadakis, A. Lavagno and P. Quarati}
\address{ Dipartimento di Fisica and INFM- 
Politecnico di Torino \\
Corso Duca degli Abruzzi 24, 10129 Torino, Italy \\ 
Istituto Nazionale di Fisica Nucleare, Sezioni di 
Cagliari e di Torino}
\maketitle
\begin {abstract} 
In this paper we will show that, because of the long-range microscopic 
memory of the random force, acting in the solar core, mainly on the electrons 
and the protons than on the light and heavy ions (or, equally, because of 
anomalous diffusion of solar core constituents
of light mass and of normal diffusion of heavy ions), 
the equilibrium statistical distribution that these particles must obey, 
is that of generalized  Boltzmann-Gibbs statistics (or the Tsallis 
non-extensive 
statistics), the distribution differing very slightly from the usual Maxwellian 
distribution. Due to the high-energy depleted tail of the distribution, 
the nuclear 
rates are reduced and, using earlier results on the standard solar model 
neutrino 
fluxes, calculated by Clayton and collaborators, we can evaluate fluxes in good 
agreement with the experimental data. While proton distribution is only very 
slightly 
different from Maxwellian there is a little more difference with electron 
distribution. 
We can define one central electron temperature as a few percent higher than the 
ion central temperature nearly equal to the standard solar model temperature. 
The difference is related to the different reductions with respect to the 
standard solar model values needed for $B$ and $CNO$ neutrinos and for $Be$ 
neutrinos. 
\end{abstract}

\date{today}

\section{Introduction}
After many years of experimental and theoretical work in different 
laboratories, 
beyond the demonstration that the sun produces energy via nuclear fusion 
reactions 
(solar neutrinos have been detected with fluxes and energies approximately 
in the 
ranges predicted by the known standard solar models (SSM), 
see Ref.s \cite{bah1,bah2,cala,caste,dar} for the up to date 
state of the art in this field), the accepted conclusions for standard solar 
neutrinos are the following. 

The solar luminosity is well known and the experimental results (Gallex, Sage, 
Chlorine, Kamiokande) seem to be mutually inconsistent; even if we discharge 
one of the 
four experiments. 

Even neglecting these inconsistencies, the measured experimental fluxes of $Be$ 
and $CNO$ neutrinos are significantly smaller than the SSM predictions (we need 
a reduction factor of about $8$ for $\Phi_{Be}$, about $2.4$ for $\Phi_{CNO}$ 
and of 2.28 for $\Phi_B$, see Table 1). 

The different reductions of the $^7Be$ and $^8B$ neutrino fluxes with respect 
to 
the SSM predictions are essentially in contradiction with the fact that both 
$^7Be$ and $^8B$ originate from the same parent $^7Be$ nuclei. 
Nevertheless, we must remember, in relation to what will be stated in the 
following paragraphs, that in the reaction 
$^7Be+e^-\rightarrow ^7Li+\nu_e$ beryllium nuclei react with the electrons; 
while in the reaction $^7Be+p \rightarrow ^8B+\gamma$, 
$^8B\rightarrow 2 \alpha+e^+ +\nu_e$, $^7 Be$ nuclei react with the protons. 

The above results could suggest that the hypothesis that nothing happens 
to the neutrinos after they are created in the interior of sun is incorrect. 
The alternative is that some of the experiments are wrong, this must be 
checked. 

There are accepted strategies to construct non-standard solar models in 
order to 
calculate reduced $^7Be$ and $^8B$ neutrino fluxes. They are as follows:\\
a) To produce models with a central temperature $T_c$ smaller than 
$T_c^{SSM}$.\\
b) To play with the nuclear cross sections determining the branches of the 
fusion chain. 
However, we must say that the resonance $^3He-^3He$ has not been found in 
the very 
recent measurements of the cross section by the LUNA group at Gran Sasso 
Laboratory \cite{arpe} 
and, as it has been recently shown by Oberh\"ummer \cite{ober}, we are 
incapable of solving, in the 
pure nuclear physics framework, the solar neutrino 
problem. \\
c) To take into account correctly the effects of screening the nuclear charges 
by the stellar plasma \cite{bra}. In order to understand its contribution 
to the fluxes and 
make a comparison with the results from the laboratory experiments a new 
theory of 
energy loss at very low energies is needed. Very recently Brown and Sawyer 
have calculated 
the screening effects in stellar plasma and have found that the enhancement 
factor to 
the fusion cross sections is very close to one \cite{bro}.\\
d) To calculate from the beginning the opacity, taking into account plasma 
collective 
effects, as done recently by Tsitovitch and collaborators \cite{tsy} and by 
Ricci \cite{ric}. 
However, we must remember that, in this case, the reaction rates must be 
recalculated 
if the electron distribution deviates from the Maxwellian one due to the 
presence of 
particular collective effects, external forces or other 
factors. 

Generally it is accepted that for standard neutrinos the actual 
experimental results seem to 
be in disagreement with SSM and with all the solar models we could build 
even considering neutrino 
oscillations \cite{kra1,kra2}.

In spite of these arguments, we wish to discuss, within the standard solar 
model description, 
the thermostatistics to which ions and electrons of the solar core should 
adhere. 
In fact, we want to show that the Maxwell-Boltzmann statistics must be 
substituted with 
the non-extensive statistics, suitable for particles subjected to long-range 
microscopic memory. \\
The resulting statistical distribution is, for the solar core, a very modest 
modification of the standard Maxwell-Boltzmann distribution \cite{ka1,qua}. 

The memory acts on the dynamics of the particles of the solar core because 
the electron mass is 
much smaller than the mass 
of the other constituents of the core plasma and because the proton 
mass is smaller than that of light and heavier ions. 
The normal memory is related to a time correlation of 
the random force which is a delta function. 
In this case the diffusion is normal, the equilibrium distribution is 
Maxwellian. 
The long-range memory implies a random force with time correlation 
different from a delta function as suggested by the results of measurements 
of neutrino fluxes. \\
The diffusion coefficient of a system of particles, subjected to these forces, 
is anomalous (non-Brownian) and the equilibrium distribution function is a 
non-extensive Tsallis distribution \cite{tsa,tsabis}. 

We describe, in the following sections, the main features of 
the non-extensive Tsallis 
statistics, the normal and anomalous diffusion coefficient, a few 
applications of 
astrophysical interest, the application to the solar neutrino production using 
the results from 
the SSM of Clayton and collaborators \cite{cla1,cla2} and give the physical 
motivations of the 
validity using, in this case, the non-extensive statistics. \\
In the Appendix A we discuss 
the anomalous diffusion coefficient in the framework of the Tsallis 
statistics because this 
quantity represents the link between the distribution adopted and the 
motivation of its use (i.e. the long-range 
time correlation of the random force). 

\section{The use of non-Maxwellian distributions in the recent past}
A customary mode of procedure to change the SSM nuclear rates can be based 
on the use 
of the appropriate statistics (the generalization of the Boltzmann-Gibbs 
statistics 
or the Tsallis non-extensive thermostatistics \cite{tsa}) which describes 
the behavior of the 
different particles composing the solar core. 
In this work we introduce a modified Maxwellian function which describes an 
equilibrium 
distribution and is the first order approximation of the Tsallis distribution 
function. \\
The physical motivations of this choice are of a many-body nature as it will be 
explained later. This choice allows a reduction of the Maxwellian rates and a 
progression towards calculations of the neutrino fluxes not in conflict with 
experimental results. 

To change the statistics and, of course, the distribution function means 
also to change the energy 
spectra of the neutrinos produced. Possibly, in future experiments at 
Gran Sasso 
Laboratory \cite{bell} or, for instance, in Hellaz experiment \cite{ypsi}, 
spectra could be measured 
with enough precision to allow the verification of the validity of the 
generalized, non-extensive statistics in the solar plasma.

In the past, after the suggestion of Kocharov \cite{kocha}, Clayton and 
collaborators \cite{cla1,cla2} 
calculated the evolutionary sequences for the sun under the assumption 
that the relative 
energies of the ions are not exactly Maxwellian.\\
Depletion of the high energy tail of the distribution 
function of relative energies in excess of 
$10$ $kT$ produces a marked reduction in the counting rate of the $Cl$ 
experiment, while producing minimal changes in the usual solar models.\\
Those changes that do occur can be compensated for by small changes in the 
initial helium concentration.\\
The mechanism leaves its own signature in the nuclear rates and in the energy 
spectra of solar neutrinos. 

Clayton could not calculate a physical cause of the deficiency 
in the number of energetic pairs and suspected that the problem lies in 
many-body physics: 
long range Coulomb interaction conspires in some unknown way to quench 
relative energies above $\approx 15$ $kT$.\\
Correction to the Maxwellian distribution function is given by the factor 
$\exp[-\delta (E/kT)^2]$. Clayton found for the solar neutrino problem 
$\delta=0.01$ ($E$ is the c.m. energy).

Departure of the two-particle distribution from the Maxwellian distribution 
function in the solar core is not unphysical. It should be possible and, 
as we will see in this 
work, the reason is long-range microscopic memory \cite{mori,qua2}, more than 
the long-range interactions (Coulomb and gravitational).

Haubold and Mathai \cite{hau,hau2}, avoiding mathematical approximations, 
derived closed form 
representations of nuclear reaction rates including resonant, non resonant, 
screened cases in terms of special functions. All these calculations have been 
accomplished without attributing an exact and precise physical meaning of the 
depletion 
of the Maxwellian distribution function used, except the following, 
that many-body effects 
in the solar core should be taken into account.\\
We also performed in the recent past calculations of nuclear rates and 
recombination cross 
sections for opacity using non-Maxwellian distributions \cite{lape,erdas}, 
before our first application of non-extensive statistics 
to solar neutrino probem \cite{ka1,qua}.

The inclusion of the helium diffusion in solar models has been widely studied 
by Bahcall 
and collaborators in the frame of Maxwell-Boltzmann statistics \cite{bah3,bah4}.

In this paper, because of physical motivations, we want to introduce, for 
the description 
of the solar plasma, a generalization of the Boltzmann-Gibbs statistics and by 
means of a reevaluation of the nuclear rates of interest to calculate the 
neutrino fluxes. 

\section{Non-extensive statistics}
Recently, non-extensive statistics (generalization of Boltzmann-Gibbs 
statistics), based 
on a new generalized definition of entropy, becoming the well known standard 
definition when 
the characteristic parameter $q\rightarrow 1$, is widely considered in many 
different physical phenomena.\\
The distribution function of these statistics has not only a depleted high 
energy tail, but in reality vanishes when:

\begin{equation}
E=\frac{k T}{1-q} \ \ \ \ \ \ {\rm if} \ \  q<1 \ \ . 
\end{equation}
Tsallis \cite{tsa} introduced in 1988 
this generalization of thermodynamics and statistical physics 
to describe systems with long-range interaction (gravitational) 
and long-range memory \cite{bogo,jund,tsa2}.\\

Let us describe briefly Tsallis statistics.\\
For a system with $W$ microscopic state probabilities $p_i \geq 0$, 
normalized as $\sum_i^W p_i=1$, we have $2$ axioms:\\
1) the entropy is 
\begin{equation}
S_q=\sum_{i=1}^W S_q^{\, i} \ \  \ \ \ \ \   
S_q^{\, i}=\frac{k}{q-1} \, p_i (1-p_i^{q-1}) \ \ ,
\end{equation}
for $q \rightarrow 1$, $S_1=-k \sum_i p_i \log p_i$. \\
2) given an observable $O$ with $o_i$ eigenvalues, the mean value is \\
$O_q=\sum_{i=1}^W p_i^q o_i, \ \ q\rightarrow 1: 
\ \ O_1=\sum_{i=1}^W p_i o_i$.\\
The validity of the two axioms lies in the comparison with experiments.\\
Properties are the following: 
systems with $q<1$ give more weight to rare events, systems with $q>1$ 
give more weight to frequent events.

The generalized entropy $S_q$ is positive, 
the microcanonical ensemble has equiprobability, 
entropy is concave ($q>0$), convex ($q<0$); Legendre transformation 
structure of thermodynamics is invariant for all $q$.\\
Additive rule (non-extensivity) is :
\begin{equation}
S_q(A \cup B)=S_q(A)+S_q (B)+(1-q) S_q (A) S_q (B)  \ \ .
\end{equation}
The Tsallis distribution function is given by the expression
\begin{eqnarray}
p_i=\frac{1}{Z_q} \left [1-(1-q) \frac{E_i}{kT} \right]^{1/(1-q)}  \ \ ,
\end{eqnarray}
and 
\begin{equation}
 Z_q=\sum_{i=1}^W \left [1-(1-q)\frac{E_i}{kT} \right ]^{1/(1-q)} \ \ .
\end{equation}

Application of generalized non-extensive statistics are the following subjects: 
matter distribution of self gravitating systems, turbulence, anomalous 
diffusion, cosmological models, big bang 
background radiation \cite{bogo,jund,tsa2,tsa3,hua,tsa4,zan,cace,zan2,pla}.
We describe here only those applications related to astrophysical problems 
(see Sections 5, 6 and 7).

In Fig.1 the behavior of the function $p_i$ is shown for values of 
$q$ greater and smaller than one.

\section{Normal and anomalous diffusion coefficients}
Is is well known that neutrinos of the four solar fluxes $\Phi_p$, $\Phi_{Be}$, 
$\Phi_{CNO}$ and $\Phi_B$ are produced in an equilibrium plasma in the solar 
core. 
In fact, the average time $\tau_{nucl}$ between two fusion reactions is 
much greater than 
the average time between two Coulomb collisions: 
$\tau_{nucl}\gg\tau_{Coulomb}$. 
We can define 
\begin{equation}
\tau_{nucl}=\frac{1}{n_n}\frac{1}{<\sigma \, v>_{nucl}}  \ \ ,
\end{equation}
where $n_n$ is the nuclear density and $<\sigma \, v>_{nucl}$ is the nuclear 
fusion rate averaged over a 
Maxwellian distribution. 

The quantity $\tau_{Coulomb}$ is related to the Brownian diffusion 
coefficient \cite{risk}
\begin{equation}
D_{Br}=\frac{k T}{m}\tau_{Coulomb} \ \ . 
\end{equation}

A factor of $10^{20}$ is in favor of equilibrium and Maxwell-Boltzmann 
distribution 
function is used to describe both electrons and ions dynamics and reaction 
rates. 
In support of the validity of the Maxwell-Boltzmann distribution, explicit 
calculations in 
which the Fokker-Planck and Boltzmann equations were solved show that the 
Maxwellian tail is filled, after few Coulomb collisions, 
when the different colliding particles 
(both electrons and ions) are all considered Brownian.

It is known that neutrino production nuclear rates, reduced in respect of 
their SSM 
values, might lead to an evaluation of the solar neutrino fluxes in agreement 
with 
the solar luminosity and not too far from the results of measurements of fluxes 
accomplished in underground experiments.

Distribution functions of equilibrium are not only Maxwellian distributions. 
If the diffusion coefficient is not Brownian 
(or normal) because the medium is dense enough 
(like the solar core) and the random force has a long-time correlation, 
then the 
equilibrium steady state is described by a distribution which is not 
Maxwellian, although it may be very close to a Maxwellian. \\
Recently, we have shown that, in the energy space, we must introduce an 
anomalous 
diffusion coefficient to solve a generalized Fokker-Planck equation, given 
by the expression \cite{ka4}: 
\begin{equation}
D^{'}=D_{Br} \left [1-(1-q) \frac{E}{kT}\right] \ \ 
\end{equation}
valid up to the energy $E=kT/(1-q)$; 
to this coefficient we can relate the anomalous collision time 
\begin{equation}
\tau^{'}=\frac{1}{\rho <\sigma v>_{Coul}} 
\left [1-(1-q)\frac{E}{kT}\right] \ \ ,
\end{equation}
where $\rho$ is the average density.\\
We are still in equilibrium conditions, but the distribution is a depleted 
Maxwellian and of non-extensive Tsallis type. 

If the ions (proton, helium, light and heavy-ions) and the electrons can be 
considered 
Brownian particles, their diffusion coefficients are normal and their 
stationary 
distributions are Maxwellian.  This description is correct also in the 
presence, in the 
dynamical equations, of a random force with a time correlation equal to a delta 
function \cite{muza,wang,com1,com2}.\\
The use of the distribution appropriate to the non-extensive ($q\ne 1$)
 thermostatistics 
has the physical meaning that the description of the dynamics of the solar core 
constituents contains anomalous, rather than normal, diffusion coefficients.\\

In the solar core we must add, in the dynamical equations of the different 
constituents, 
a random force $F(t)$ with time correlation different from the delta function 
(see the Appendix A for the needed details):
\begin{equation}
<F(0)\,F(t)>=F_0(\beta) t^{-\beta} \ \ ,
\end{equation}
where $\beta$ is a parameter, in this case, smaller than one and related to 
$q$ by 
\begin{equation}
\beta=\frac{1}{2-q}=\frac{1}{1+2\delta} \ \ ,
\end{equation}
(for $\beta=1$, or $q=1$ and $\delta=0$ the delta function time correlation is 
recovered).\\
The choice of a non-delta time correlation is suggested by the comparison of 
calculations with the experimental results, i.e. 
by the values of the parameter $q$ required by 
the results of the measurements and, consequently, by the values of the 
parameter $\beta$.\\
In the solar core, 
the anomaly consist in a subdiffusion, because the value of the parameter 
$q$ to be used lies in the range: $q=0.952 \div 0.994$ 
(or $\delta=0.003\div 0.024$), 
as explained in the following Section 7. \\
The subdiffusion correction to the normal behavior is very small but 
sufficient to 
calculate neutrino fluxes in agreement with the measured fluxes. 

In the solar core, the anomaly is not due to the long range gravitational 
interaction, 
rather it is due to the long-range memory of the random force.

In the Appendix A we discuss the meaning of a 
non-linear or non-Brownian diffusion 
coefficient and the solutions of the Fokker-Planck equation appropriate to 
the non-extensive thermostatistics.

\section{Stellar polytropes}
This is the first example of application of non-extensive statistics.\\
The equation of state appropriate to the stellar polytropes is 
\begin{equation}
{\cal P}={\cal K} \rho^\gamma \ \ ,
\end{equation}
where ${\cal P}$ is the pressure, $\rho$ the density, $\gamma$ a constant 
related to 
specific heats defined by $\gamma=1+1/n$ ($n$ is the polytropic index).\\
The idrostatic equilibrium equation has spherical symmetric solutions 
corresponding to 
compact spherical configurations of self-gravitating mass (stellar polytropes).

Relation for stellar polytropes between $n$ and the Tsallis parameter $q$ 
is \cite{pla}
\begin{equation}
n=\frac{3}{2}+\frac{q}{1-q} \ \ ,
\end{equation}
The index $n$ must exceed $1/2$ to avoid singularity in the gamma function; 
$n>5$ 
gives rise to unnormalizable mass distribution (unphysical). The Tsallis 
distribution 
function has a spatial cutoff of the mass distribution (compact nature of 
the stellar polytrope). 

In the solar core we have a constraint which fixes differently the relation 
between $n$ 
and $q$ \cite{ka1,qua} 
\begin{equation}
q=1-\frac{\tau}{n+1}, \ \ \ \ \tau=\frac{k T \rho}{{\cal P}}, \ \ \ 
q=1-2\delta \ \ ,
\end{equation}
the quantity $\tau$ is very close to one if an ideal gas of particles having 
a certain 
average molecular weight is considered; in fact, the equation of state of an 
ideal gas, 
within the Tsallis statistics, 
differs from the classical one for the addition of terms whose contribution 
depends on powers of $\delta$ (in the limit $q\rightarrow 1$, 
this contribution can safely be disregarded) \cite{tsaeq}.

\section{Application to galaxy cluster velocity}
We have shown that the observational data recently provided by Giovannelli 
et al. \cite{giova}
(COBE) and discussed by Neta Bahcall and Peng Oh \cite{neta} and 
by Moscardini et al. \cite{mos} concerning 
the velocity distribution of clusters of galaxies can be naturally fitted by 
a statistical 
distribution which generalizes the Maxwell-Boltzmann one \cite{lava}.\\

In this generalization of Boltzmann Gibbs statistics the probability 
function of having 
a cluster with velocity greater than $v$ can be written as 
\begin{eqnarray}
P(>v) & = & \frac{\displaystyle{ \int_{v}^{v_{max}} dv \left[ 1-(1-q)(v/v_0)^2 
\right]^{q/(1-q)}}}{\displaystyle{ \int_{0}^{v_{max}} dv 
\left[ 1-(1-q)(v/v_0)^2 
\right]^{q/(1-q)}}}
\, \, , \nonumber 
\end{eqnarray}
\begin{eqnarray}
v_{max} & = & \left\{
\begin{array}{ll}
v_0 (1-q)^{-1/2} & \mbox{\, \, \, \, if $q<1$}   \\
\infty & \mbox{\, \, \, \, if $q \geq 1$.} 
\end{array}
\right. \, \,  , \nonumber
\end{eqnarray}
A remarkably good fitting with the data is obtained for 
$q=0.23\pm^{0.07}_{0.05}$ and $v_0=490\pm 5$ km/s.

\section{Application to solar neutrino problem}
Taking advantage of the calculations accomplished by Clayton and 
collaborators,  
based on their solar model (standard) which uses a distribution 
function of 
Maxwellian type with a depleted tail (this is a first and suitable 
approximation of the 
Tsallis distribution), we discuss the application of the 
non-extensive statistics to 
the solar neutrino problem. In Refs \cite{ka1,qua} we report 
how the constraints imposed by the solar core compel us to 
derive the relation (14) 
instead of Eq.(13) suitable for stellar polytropes; results of 
the application of the Tsallis 
statistics to solar neutrino problem are shown without mentioning 
the physical motivations 
to use this new statistics (mainly due to the anomalous diffusion 
of light particles rather 
than normal diffusion appropriate to heavier ions).

A value of the Clayton parameter $\delta$ different from zero 
(we remember that 
$q=1-2\delta$) makes the star more luminous and reduces the rate 
of energy production 
at a given temperature. The solar core contracts to higher 
temperature. However, 
the increase of temperature at given solar luminosity does not 
increase the neutrino fluxes 
that decrease with $\delta$ (except $\Phi_p$, but its growth 
is very slow and 
its value is almost constant) because of the depleted 
Maxwellian tail and the consequent 
rearrangement of the particle distribution.\\
We introduce the reduction factor $f_{red}$ of the $B$, 
$CNO$ and $Be$ fluxes as a 
function of $\delta$ (or $q$) 
\begin{equation}
\Phi_{B,CNO,Be}=\frac{1}{f_{red}}  \Phi_{B,CNO,Be}^{SSM} \ \ ,
\end{equation}
and the enhancement factor $f_{incr}$ of the proton flux 
$\Phi_p$ as a function of 
$\delta$ (or $q$) 
\begin{equation}
\Phi_{p}=f_{incr}  \Phi_{p}^{SSM} \ \ .
\end{equation}

In Fig. 2 we show both the reduction factor $f_{red}$ and 
the increasing factor 
$f_{incr}$ as function of $\delta$.\\
These two quantities are derived from the calculation of 
the neutrino fluxes as function of 
$\delta$ by Clayton and collaborators. The SSM fluxes are 
all normalized to one at 
$\delta=0$. \\
We have obtained an expression for the factors valid up to 
$\delta=0.025$ 
(an extrapolation of the results reported by Clayton and 
collaborators, in the range between 
$\delta=0.020$ and $\delta=0.025$, has been done):
\begin{eqnarray}
&&f_{red}(B)=1+92581.75 \delta^{1.92} \ \ , \nonumber \\
&&f_{red}(CNO)=1+23.6 \delta^{0.49}+30 \delta \ \ , \nonumber \\
&&f_{red}(Be)=1+851.71 \delta^{1.29} \ \ , \\
&&f_{incr}(p)=1+0.89 \delta^{0.62} \ \ . \nonumber
\end{eqnarray}

We can clearly see that the selection of $\delta=0.003$ 
($q=0.994$) produces the 
correct reduction and enhancement factors (see Section 1) 
of the SSM values of $\Phi_B$, $\Phi_{CNO}$ 
and $\Phi_p$ and then neutrino fluxes in agreement with the 
measured values and with the solar 
luminosity; the choice $\delta=0.024$ ($q=0.952$) is the 
right selection for $\Phi_{Be}$. \\
The reduction factors $f_{red}$ have been derived in such a 
way that the four fluxes 
satisfy the constraints imposed by the solar luminosity and 
by the results of 
measurements of the Gallex, Sage, Chlorine and Kamiokande 
experiments, within their 
precision \cite{cala,caste}.

The need for two different choices of $\delta$ is due to the 
$Be$ flux and depends on the electron 
distribution, while the $B$ flux (and, of course, the proton 
and $CNO$ fluxes) 
depends on the proton distribution. \\
In fact: \\
$\Phi_{Be}\propto N(e^-) N(^7 Be) <e^{-} \,\, ^{7} Be>$ and 
$\Phi_{B}\propto N(p) N(^7 Be) <p\,\, ^7 Be>$, \\
where $N(e^-)$ is the electron density, $N(p)$ the proton density, 
$<e^{-}\,\, ^{7} Be>$ 
the nuclear capture rate electron-beryllium and $<p \,\, ^7 Be>$ 
the nuclear rate proton-
beryllium (analogously we can say of the other two fluxes). 
See the Table where the results on the fluxes are reported.\\ 
The consequences of this derivation is that, in the solar core, 
electrons behave differently 
from protons and from the other light and heavy ions. The latter 
are nearly 
Maxwellian, the electron distribution must have a more depleted 
tail. In addition, 
the polytropic index $n$ of electrons differs from that of the 
other constituents. The 
solar core plasma model, consisting of two main components 
(electrons and ions), 
with different features, is under study. 
Here we still want to examine the effect of these physical 
facts on the 
central solar temperature by using the well known dependencies 
of the fluxes 
on central temperature as given in the literature \cite{caste,bah3,bah4}.

\section{Temperature of the central core}
We derive the dependence of the four fluxes upon central temperature 
$T_c$ when all SSM calculations are executed at $T^{SSM}_c$, 
taking advantage of the expressions reported in Ref. \cite{caste,bah4}. 
We have derived from 
Clayton and collaborators calculations the expression
\begin{equation}
 T_c=T_c^{SSM} (1+3.12 \delta) \ \ ,
\end{equation}
valid for $\delta<0.1$.\\
The dependence is as follows
\begin{eqnarray}
&&\Phi_{p} \propto (1-0.88 t^{11}) (1+0.44 t^{0.62}) \ \ , \nonumber \\
&&\Phi_{B} \propto T^{25}_c [1+10^4 (t-1)^{1.92}]^{-1} \ \ , \nonumber \\
&&\Phi_{CNO} \propto  (x T^{20}_c+(1-x) T_c^{23}) 
[1+13.66 (t-1)^{0.43}]^{-1} \ \ , \\
&&\Phi_{Be} \propto T^{11}_c [1+196 (t-1)^{1.29}]^{-1} \ \ , \nonumber 
\end{eqnarray}
where $t=T_c/T_c^{SSM}=1+3.12\, \delta$, $x$ represents the percentage 
of the contribution to the 
CNO flux from the ions $^{13}N$ and $^{15}O$ and $1-x$ the 
contribution from 
$^{17}F$.\\
We can see that $T_c$ increases by $1$ per cent in respect to 
$T_c^{SSM}$ using 
the value of $\delta$ appropriate to $\Phi_{p,B,CNO}$ fluxes 
(this temperature is the ionic 
plasma temperature $T_c^{(i)}=1.01 \,\, T_c^{SSM}$), and 
increases by seven per cent using 
the value of $\delta$ appropriate to $\Phi_{Be}$ (this 
temperature is the electron plasma temperature 
$T_c^{(e)}=1.07 \,\, T_c^{SSM}$).

The meaning of the results of the above relations is different 
from that of the results 
coming from the relations of other authors. While, according to 
these authors, the 
central temperature must decrease in respect to the SSM value to 
reproduce the behavior of the ratio of the fluxes, 
in our approach the central temperature $T_c$ must slightly 
increase with $\delta$ 
and the fluxes (except $\Phi_{p}$ which is almost constant) 
decrease because of the depleted 
tail and the consequent rearrangement of the particles.

Helioseismology is probing the interior structure and dynamics 
of the sun with great 
precision \cite{deru,chr,har}. 
Acoustic waves and internal gravity waves are strongly influenced 
by the central 
temperature and by the structure of central regions. 
The reliability of the 
non-extensive statistics applied to the solar core can be tested 
in the context of helioseismological studies. 
This is the topic of our investigation in the near future. 

\section{Conclusion}
The validity of the description of the solar core shown in this work, 
by means of a generalized Boltzmann-Gibbs statistics, 
could be verified in future planned experiments 
by measuring the electron central temperature $T_c$, through 
the detection of the $Be$ 
line and by measuring the neutrino energy spectra, looking for 
a shift of the Gamow peak and of the neutrino 
maximum energy due to non-extensive thermostatistics effects, 
if very high precision will be obtained.\\
It could be of practical use, from a theoretical point of view, 
to evaluate the effect of the neutrino oscillations of the 
fluxes in the context of the non-extensive statistical description 
of the solar core.

We have shown that anomalous diffusion (different from Brownian) of 
electrons and of 
protons, caused by the long-range microscopic memory of the random 
force in 
the solar core plasma, is the cause of the stationary distributions 
of the non-extensive 
statistical type we must use. 
Light and heavy ions have a normal diffusive behavior. 
As a consequence, the values of the nuclear rates of the different 
reactions decrease in respect to Maxwellian evaluations (we have 
taken advantage of the 
SSM of Clayton and collaborators). 

A particular choice of the value of the non-extensive Tsallis 
parameter $q$ for the 
electrons ($q=0.952$) and for the protons ($q=0.994$) allows the 
calculation of neutrino 
fluxes. The values of $q$ differ from the value $q=1$ (Maxwellian 
distribution) for very small 
magnitudes. The value $q=1$ is assumed for light and heavy ions. \\ 
This description implies that the electrons have a behavior which 
differs from the Maxwellian behavior more than that of the protons, 
while we can assume that light 
and heavy ions have a normal behavior, or normal diffusion coefficient, 
which implies 
Maxwellian distribution.

Our results, based on the standard solar model developed by Clayton and 
collaborators, 
imply a weak increase of the central temperatures (both electronic 
and ionic), the 
fluxes decrease because of the depleted tail of the distribution. 

Finally, we must remember that the validity of the non-extensive 
statistics in the solar core plasma is also based 
on the polytropic nature of the sun, whose polytropic index $n$ 
has a finite value with the consequence that 
the distribution must generally be non-Maxwellian. \\
The approach described in this work will be refined in order to 
take into account the information 
given by the helioseismological studies and measurements. \\
Our approach is based on a standard solar model with diffusion 
(anomalous and normal) 
of the constituents of the solar core plasma, 
therefore it is highly possible that the test with 
the helioseismological data be positive, but it must be checked; 
this will be accomplished 
soon.

We realize that the results reported in this paper are not definitive. 
In fact we do not 
indicate errors and precisions of the figures we have derived and shown. 
This is due to the reason that we do not have used a solar model 
code and we have not accomplished calculations abinitio. 
However, we have taken advantage of a standard solar model previsions 
and our results are an indication of a possible solution of the solar 
neutrino problem 
(anomalous diffusion for the lighter constituents of the core). 
We hope that in the next future the non-extensive Tsallis statistics 
will be tested within one 
of the standard solar models actually in use.

\section{Appendix A}
In the solar core we can distinguish the following constituents:\\
protons; helium ions; other light ions and heavy ions 
and electrons (with $m_{ion}\gg m_{He}>m_p\gg m_e$).\\
Due to the magnitude of $m_e$, the electrons diffusive behavior 
is a little less normal than 
that of light ions, because their dynamics, described for instance 
by a generalized 
Langevin equation, require the inclusion of a particular random 
force. For the 
same reason, the magnitude of their mass, heavy ions have a normal 
diffusive behavior.
We assume that also helium has a normal diffusion.\\
More explicitly, the deterministic equation $m \dot{v}+\alpha v=0$, 
whose solution is 
$v(t)=v(0) \exp(-\gamma t)$ ($\gamma=\alpha/m=\tau^{-1}$), 
is not sufficient to 
describe the diffusion of particles with a mass smaller than 
the masses of the particles 
composing the rest of the plasma (their velocity due to thermal 
fluctuations is considerable). \\
If the time correlation of the random force $F(t)$, which 
we must add, is a delta function, 
the equilibrium distribution is also a Maxwellian distribution.\\
From another aspect, a long-range microscopic memory, i.e. a 
time correlation of the random 
force different from a delta function, produces an equilibrium 
distribution 
of a non-extensive statistical category.\\
Subdiffusion is responsible for the depletion of the Maxwellian tail. 
The equilibrium 
distribution of electrons and of protons and helium 
ions can be derived from 
dynamical equations containing a colored noise, more 
coloured for electrons than for 
proton and for helium ions. 

Anomalous diffusion, like subdiffusion, is related to the 
normalized equilibrium velocity 
autocorrelation function $C_v (t)$ \cite{muza}
\begin{equation}
C_v(t)=\frac{1}{<v^2>} <v(0) v(t)> \ \ ,
\end{equation}
where $v(t)$ is the velocity of the particle at time $t$ and 
$C_v(t)$ does non depend on the 
time origin.\\
Mean square displacement is 
\begin{equation}
<x(t)^2>=t \int^t_0 C_v(\tau) d\tau - \int^t_0 \tau C_v(\tau) d\tau \ \ .
\end{equation}
If $C_v(u)$ decays faster than $u^{-2}$ after a long period of 
time, one observes 
normal (or Brownian) diffusion.\\
We are interested here in subdiffusion because our analysis of 
the measured neutrino fluxes 
suggests that this is the type of diffusion of the solar core particles. 
This regime is reached when 
$C_v(u)\approx A u^{-(2-\beta)}$ as $u\rightarrow\infty$ 
($0<\beta<1$, $A<0$ and 
$\int_0^{\infty} C_v(u) du=0)$ then 
$<x(t)^2>\approx B t^\beta$ as $t\rightarrow\infty$.\\
If the particle is moving in the positive $x$ direction, 
it is more likely to move in the 
negative $x$ direction in the next instant. The fluctuating 
velocity reverses its 
direction very often.

The time correlation of the random force can be written, for 
instance, as 
\begin{equation}
<F(0) F(t)>=\frac{D}{\vartheta} \exp(-t/\vartheta) \ \ .
\end{equation}
If $\vartheta\rightarrow 0$ the correlation is 
given by $2\, D\delta(t)$ which refers 
to a normal diffusion (therefore $q$ is linked to $\vartheta$).

Following Wang \cite{wang}, we can write generally
\begin{equation}
<F(0) F(t)>=C_f(t)=F_0 (\beta) t^{-\beta} \ \ .
\end{equation}

We relate now the parameter $\beta$ to the Tsallis parameter $q$. 
Tsallis and Bukmann \cite{tsa4} 
have shown that correlated anomalous diffusions are described by 
the following generalized 
Fokker-Planck equation in the space coordinate \cite{tsa4} 
\begin{equation}
\frac{\partial}{\partial t}[p(x,t)]^\mu=-\frac{\partial}{\partial x} 
\left \{ {\cal F}(x) [p(x,t)]^\mu \right \} +
D \frac{\partial^2}{\partial x^2}
[p(x,t)]^\nu \ \ ,
\end{equation}
where ${\cal F}(x)=-\partial V/\partial x$ is an external (drift) force 
associated with the potential 
$V(x)$ (${\cal F}(x)=k_1-k_2 x$, $k_2 \ge 0$). \\
The Tsallis parameter $q$ is related to $\mu$ and $\nu$ 
\begin{equation}
q=1+\mu-\nu \ \ .
\end{equation}
As here we want to preserve the norm of the distribution, 
we must pose $\mu=1$.\\
Therefore: $q=2-\nu=1-2\delta$ and then $\nu=1+2\delta$ 
($\delta$ is the Clayton 
parameter). By comparison, we have the relation 
\begin{equation}
\beta=\frac{1}{2-q}=\frac{1}{1+2\delta}=\frac{1}{\nu} \ \ .
\end{equation}
The normal diffusion is recovered when $\beta=1$ ($\nu=1$), 
i.e., as we know, when 
$q=1$ ($\delta=0$).

The Fokker Plank equation for the particles of the solar core is ($\mu=1$, 
$\nu=1-2\delta$)
\begin{equation}
\frac{\partial}{\partial t} p(x,t)=-\frac{\partial}{\partial x} 
\left \{ {\cal F}(x) p(x,t) \right \} +D \frac{\partial^2}{\partial x^2}
[p(x,t)]^{1+2\delta} \ \ .
\end{equation}
The stationary solution of the above equation is the Tsallis 
distribution in the coordinate 
space ($x$ has the dimension of an energy square root)
\begin{eqnarray}
p(x)=\left\{\begin{array}{ll}
{\cal N}(q,kT) \left [1-\frac{(1-q) x^2}{kT}\right ]^{1/(1-q)}, 
& \mbox{$\mid x\mid <\sqrt{\frac{k T}{1-q}}$}   \\
0 & \mbox{otherwise} 
\end{array}
\right. 
\end{eqnarray}
where ${\cal N}$ is the following normalization constant

\begin{equation}
{\cal N}(q,kT)=\pi^{-1/2} \left (\frac{kT}{1-q}\right )^{-1/2} \, \, 
\frac{\displaystyle{\Gamma\left(\frac{5-3 q}{2 (1-q)}\right)}}
{\displaystyle{\Gamma \left (\frac{2-q}{1-q}\right)}} \ \ .
\end{equation}

The results obtained in the Appendix can be recovered using the 
approach described in 
Ref.\cite{ka4} where a new definition of the diffusion coefficient 
in the momentum space, 
whose expression is reported in Eq.(8) of the text, was introduced.

In the momentum space the equilibrium distribution can be derived 
by means of the Fourier 
transformation of the distribution $p(x)$, following the procedure 
outlined by Tsallis 
et al. \cite{tsa4} (its analytic expression is in terms of the 
modified Bessel function ${\cal K}$ \cite{mathai}). 
In the situations where $q\rightarrow 1$ (as in the case of the solar 
neutrino problem) 
we can easily verify that the Tsallis distribution in the coordinate space, 
given by Eq.(28), corresponds 
to another Tsallis distribution in the momentum space. In fact, 
for $q\rightarrow 1$ we 
can write (here $x$ is adimensional):
\begin{equation}
p(x)\propto [1-(1-q) x^2]^{1/(1-q)}\simeq \exp(-x^2-\frac{1-q}{2} x^4) \ \ .
\end{equation}
The Fourier transformation of $p(x)$ is the Clayton distribution 
\begin{equation}
{\cal F}[p(x)]\simeq \exp^{-\frac{{\rm p}^2}{4}} 
[1+(1-q)\frac{{\rm p}^2}{4}-(1-q)\frac{{\rm p}^4}{32}]
\simeq \exp[-\frac{{\rm p}^2}{4}-(1-q)\frac{{\rm p}^4}{32}]\ \ ,
\end{equation}
which approximates, in the momentum space, the Tsallis distribution
\begin{equation}
{\cal F}[p(x)]\propto [1-(1-q) {\rm p}^2/4]^{1/(1-q)}\ \ .
\end{equation}

\vspace{2cm}

{\bf Figure Caption}\\
\\
{\bf Fig.1}\\
The Tsallis distribution function $p(E_i)$ for different values of $q$: $q=0$, 
$0.5$, $1$ (MB), $1.5$. The typical depletion of the high-energy 
tail of the Maxwellian 
($q=1$) is clearly evidentiated for $q<1$.\\
\\
\\
\noindent
{\bf Fig.2}\\
The reduction factor $f_{red}$ ($B$, $CNO$, $Be$) and the 
increasing factor $f_{incr}(p)$ 
as functions of the parameter $\delta$ ($q=1-2\delta$), up to $\delta=0.025$. 
We call the attention to the fact that SSM values of the $B$, 
$CNO$ and $Be$ fluxes must be divided by 
$f_{red}$ ($B$, $CNO$, $Be$) to obtain the fluxes within the non-extensive 
statistics, while the SSM proton flux must be multiplied by $f_{incr}(p)$. 
All SSM fluxes are normalized to one at $\delta=0$.
\\
\\
\\
\\
\\
\\

\noindent
{\bf Table Caption}\\
\\
{\bf Table 1}\\
The SSM fluxes ($\delta=0$) from Ref.\cite{bah4}, the flux
$\Phi_p=f_{incr}\Phi_p^{SSM}$ and the fluxes 
$\Phi_{B,CNO,Be}=\Phi_{B,CNO,Be}^{SSM}/f_{red}$ evaluated within 
the non-extensive statistics. 
All fluxes are in $10^9$ cm$^{-2}$ s$^{-1}$ but $\Phi_{B}$ in 
$10^6$ cm$^{-2}$ s$^{-1}$
(we recall that the Kamiokande experiment 
gives the result $\Phi_B=2.9\pm 0.4$).

$$\offinterlineskip \tabskip=0pt 
\vbox{
\halign 
{\strut  \vrule\vrule#& \quad \hfil # \hfil \quad &
\vrule#& \quad \hfil # \hfil &
\vrule#& \quad \hfil # \hfil &
\vrule#& \quad \hfil # \hfil &
\vrule#& \quad \hfil # \hfil &
\vrule\vrule# \cr
\noalign {\hrule}
\noalign {\hrule}
&$\delta$ && $\Phi_p$ && $\Phi_{B}$ && $\Phi_{CNO}$ && $\Phi_{Be}$ & \cr
\noalign {\hrule}
\noalign {\hrule}
& 0 && 59.24(1$\pm$ 0.01) && 6.62(1$^{+0.14}_{-0.17}$) && 1.17(1$\pm$0.2) 
&& 5.15(1$\pm$0.06) &\cr
\noalign {\hrule}
& 0.003 && 60.66 ($f_{incr}$=1.024) && 2.85 ($f_{red}$=2.326) && 0.48 
($f_{red}$=2.46) &&   &\cr
\noalign {\hrule}
& 0.024 &&  &&  &&  && 0.65 ($f_{red}$=7.93) &\cr
\noalign {\hrule}
\noalign {\hrule} }}$$

\end{document}